\documentclass[preprint,12pt]{elsarticle}

\usepackage{graphicx}
\usepackage{subfigure}
\usepackage{amsmath}

\usepackage{amssymb}
\usepackage{graphicx}

\journal{Physica Medica}

\begin{document}

\begin{frontmatter}

\title{High frequency residues: a new set of signals for detectability studies of an X-ray imaging system}

\author{Antonio Gonz\'alez-L\'opez\corref{cor1}}
\ead{antonio.gonzalez7@carm.es}
\cortext[cor1]{Corresponding author}
\address{Hospital Universitario Virgen de la Arrixaca, ctra. Madrid-Cartagena s/n, 30120 El Palmar (Murcia), Spain}

\begin{abstract}
A new set of signals for studying detectability of an x-ray imaging system is presented. The results obtained with these signals are intended to complement the NEQ results.\\
The signals are generated from line spread profiles by progressively removing their lower frequency components and the resulting high frequency residues (HFRs) form the set of signals to be used in detectability studies. Detectability indexes for these HFRs are obtained using a non-prewhitening (NPW) observer and a series of edge images are used to obtain the HFRs, the covariance matrices required by the NPW model and the MTF and NPS used in NEQ calculations. The template used in the model is obtained by simulating the processes of blurring and sampling of the edge images. Comparison between detectability indexes for the HFRs and NEQ are carried out for different acquisition techniques using different beam qualities and doses.\\
The relative sensitivity shown by detectability indexes using HFRs is higher than that of NEQ, especially at lower doses. Also, the different observers produce different results at high doses: while the ideal Bayesian observer used by NEQ distinguishes between beam qualities, the NPW used with the HFRs produces no differences between them.\\
Delta functions used in HFR are the opposite of complex exponential functions in terms of their support in the spatial and frequency domains. Since NEQ can be interpreted as detectability of these complex exponential functions, detectability of HFRs is presented as a natural complement to NEQ in the performance assessment of an imaging system.
\end{abstract}


\end{frontmatter}

%
%
\section{Introduction}
Quality control tests of x-ray imaging devices are carried out as part of diagnostic imaging quality reference programs. They are carried out during the system acceptance and the commissioning of the imaging systems and are part of periodical quality assurance programs \cite{samei18,marshall11}. Their objectives are to allow prompt corrective action to maintain x-ray image quality and to reach an optimized performance in both image quality and dose \cite{yan12,lin15,delis17,jang18}.

Image quality has been traditionally assessed by determining physical parameters of the imaging sensor such as spatial resolution and noise \cite{strudley15,solomon15,gonzalez20a,gonzalez20b}. In addition, operational performance carried out by means of task-based tests has also been and is currently being developed  \cite{samei19,hernandez15,eck15,peteghem16,evan20}. The purpose of these tests is to determine the clinical performance of the imaging system, in such a way that the evaluation results are more consistent with its diagnostic capabilities.

According to Tapiovaara and Wagner \cite{tapiovaara93}, one of the fundamental imaging stages (image data acquisition) can be analyzed rigorously by using the concept of the ideal Bayesian observer (IBO). In this way, the intrinsic detectability performance of the detector can be evaluated without the human observer intervention. In addition to the ideal observer, the non-prewhitening (NPW) matching filter is also often used for assessing detectability tasks in imaging systems. In fact, the NPW has shown to reproduce the performance of a human observer when combined with human eye response (NPWE) \cite{ICRU1995,1388565,Richard2008,10.1117/12.2081655,BOUWMAN20161559}.

An example of how the ideal Bayesian observer describes detectability of a system can be found in the noise equivalent quanta (NEQ), since the NEQ at a given frequency gives the detectability of a harmonic signal with that frequency. However, the sinusoidal-type signals for which NEQ provides detectability are very different from the signals generally sought after in diagnostic radiology. A more realistic set of signals would have compact support and would not be as regular as a sine wave is. For this reason, studying detectability using other type of signals could contribute to a better understanding of the system performance.

In a previous work \cite{gonzalez21}, wavelet analysis has been used in combination with a star-bar object to generate a set of signals with different frequency content. Then, detectability for the set of signals was calculated and its dependency on the frequency was presented. In this way, an alternative analysis to that of NEQ was carried out using a more representative set of signals. 

In this work, a similar approach to that of the aforementioned work \cite{gonzalez21} is carried out and image data acquisition is studied using beams with different qualities and dose levels. For the analysis, images of an edge phantom are used to obtain different test signals in which detectability is investigated using the NPW observer. These test signals are the output of linear transformations of the image and can be seen as high frequency residues of line profiles.

\section{Material and methods}
\subsection{Theory}\label{sec:theory}
The main goal of this sub-section is to illustrate, in a particular case, how NEQ describes detectability in a discrete system. Throughout this subsection a one dimensional, discrete finite, shift invariant and linear system will be considered. Inputs, outputs and noise will be periodic as, for instance, signals arising from circular scans in a star-bar phantom \cite{gonzalez20a}. Also the sampling distance will be $\Delta=1$ and noise will be assumed to be additive, Gaussian and wide sense stationary. In such a system the output $g=\{g_k\}_{k=0..N-1}$ can be described from the input $f$, the system transfer matrix $H$ and noise $n$ as $g=Hf+n$, and the detectability index $d^2$ expressing the squared signal to noise ratio (SNR) of the decision function of an IBO in an SKE/BKE task can be calculated as \cite{ICRU1995}
\begin{equation}\label{eq:snr2}
d^2_{IBO}=(Hf)^tC^{-1}_nHf,
\end{equation}
where $C_n$ is the covariance matrix of noise. This detectability index can also be obtained as (see appendix \ref{ap:1})
\begin{equation}\label{eq:d2ibofrec}
d^2_{IBO}=\sum_{u=0}^{N-1}\frac{|\hat{f}_u|^2 |\hat{h}_u|^2}{\hat{c}_u},
\end{equation}
where $\{\hat{f_u}\}_{u=0..N-1}$ is the discrete Fourier transform (DFT) of input $f$, and $\{\hat{h_u}\}_{u=0..N-1}$ and $\{\hat{c_u}\}_{u=0..N-1}$ are the DFTs of the first rows of $H$ and $C_n$ respectively. It should be noted that under the periodicity assumption $\{\hat{h_u}\}$ and $\{\hat{c_u}\}$ are the optical transfer function of the system and the noise power spectrum respectively.

Equation \ref{eq:d2ibofrec} shows that if a complex exponential function $f_v=\{Ke^{i2\pi kv}/N\}_{k=0..N-1}$ is used as the input, then $d^2_{IBO}$ will become the NEQ value at frequency $v$ in equation \ref{eq:neq} (note that the DFT of $f_v$ is $K\delta_{uv}$, $\delta_{uv}$ being the Kronecker delta). In this way, the NEQ value at a given frequency $v$ can be interpreted as the detectability of the harmonic signal that spans the entire signal domain and is characterized by that frequency,
\begin{equation}\label{eq:neq}
NEQ_u=\frac{|\hat{h}_u|^2}{\hat{c}_u/|K|^2},\:u=0..N-1.
\end{equation}

Therefore, given a radiation quality and an entrance air KERMA, the NEQ provides information on how detectability is lost as frequency increases. This relationship provides a comprehensive description of the system and can be used for describing detectability of any other input signal $f$ (equations \ref{eq:d2ibofrec} and \ref{eq:neq}). However, as mentioned above, sinusoidal-type signals are poorly representative of signals found in diagnostic radiology.

This work analyses the system detectability using a particular set of signals characterized by the lower bound of their frequency content. In this way, a parallel analysis to that of the NEQ-frequency curve is performed. Instead of using an IBO observer, employed in this subsection to show an example of how detectability is described by NEQ, a NPW observer will be used. For this observer, the squared SNR $d^2_{NPW}$ of its decision function is \cite{ICRU1995}
\begin{equation}\label{eq:d2npw}
d^2_{NPW}=\frac{((Hf)^tHf)^2}{(Hf)^tC_gHf},
\end{equation}
where $C_g$ is the average of the covariance matrices for $g$ and $n$.
\subsection{Images and signals}
The x ray beam was generated in an Ysio Max X-ray system from Siemens. The beam qualities used were RQA 3 and RQA 5. Source-detector distance was set to $153$ cm and the field size was set to 15.8 cm $\times$ 15.4 cm. 

For each radiation quality two entrance air KERMA were used, 1.79 and 17.9 $\mu$Gy for RQA 3 and 1.76 and 17.6 $\mu$Gy for RQA 5. The imaging detector was a PIXIUM 3543 pR from Trixell (CsI coupled to TFT matrix of $a$Si with a pixel spacing of 0.144 mm $\times$ 0.144 mm). 

\begin{figure}
\centering
	\includegraphics[width=80mm]{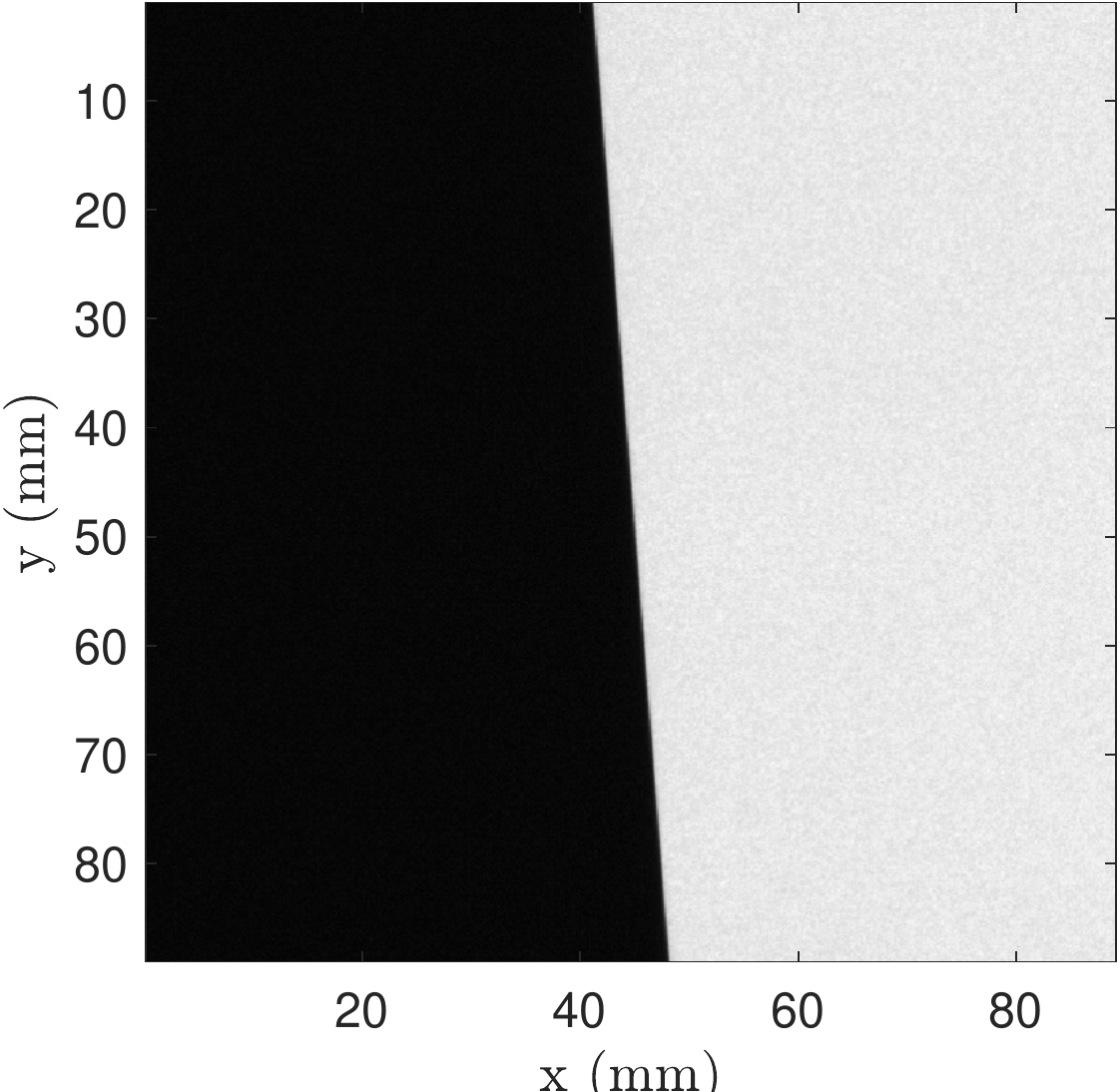}					
	\caption{Central part of the image of an edge phantom acquired with a beam quality RQA 3 and an entrance air KERMA of 1.79 $\mu Gy$.}
	\label{fig:edge}
\end{figure}

The edge phantom was a TX5 (Scanditronix-Welhöfer). It has a tungsten edge $1$ mm thick and 100 mm length. A central part of the image of 90 mm $\times$ 90 mm was used for calculation (figure \ref{fig:edge}).

\begin{figure}
	\centering
	\includegraphics[width=140mm]{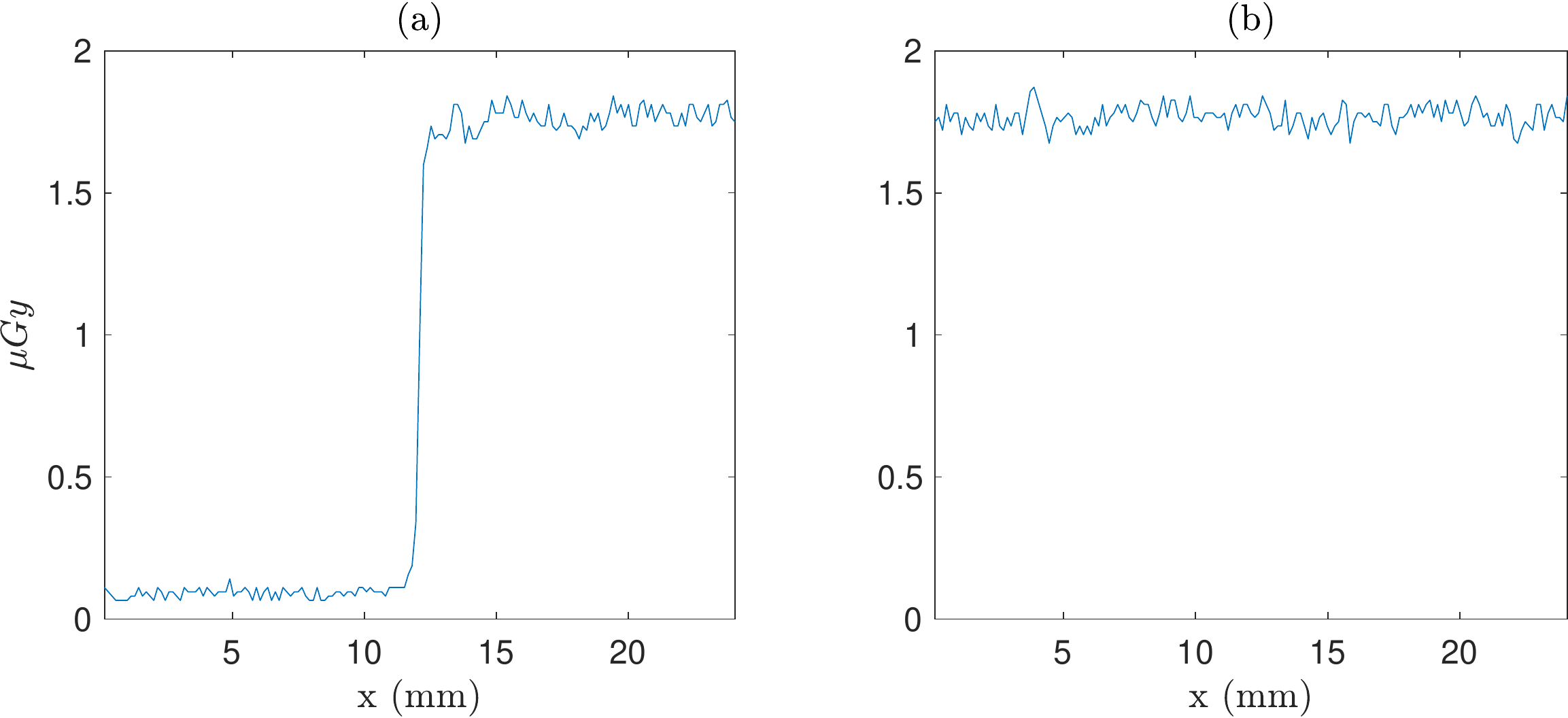}
\caption{Central (a) and right (b) parts of the profile resulting from scanning the edge image in figure \ref{fig:edge} along the top row.}
\label{fig:esf}
\end{figure}

\begin{figure}
	\centering
	\includegraphics[width=140mm]{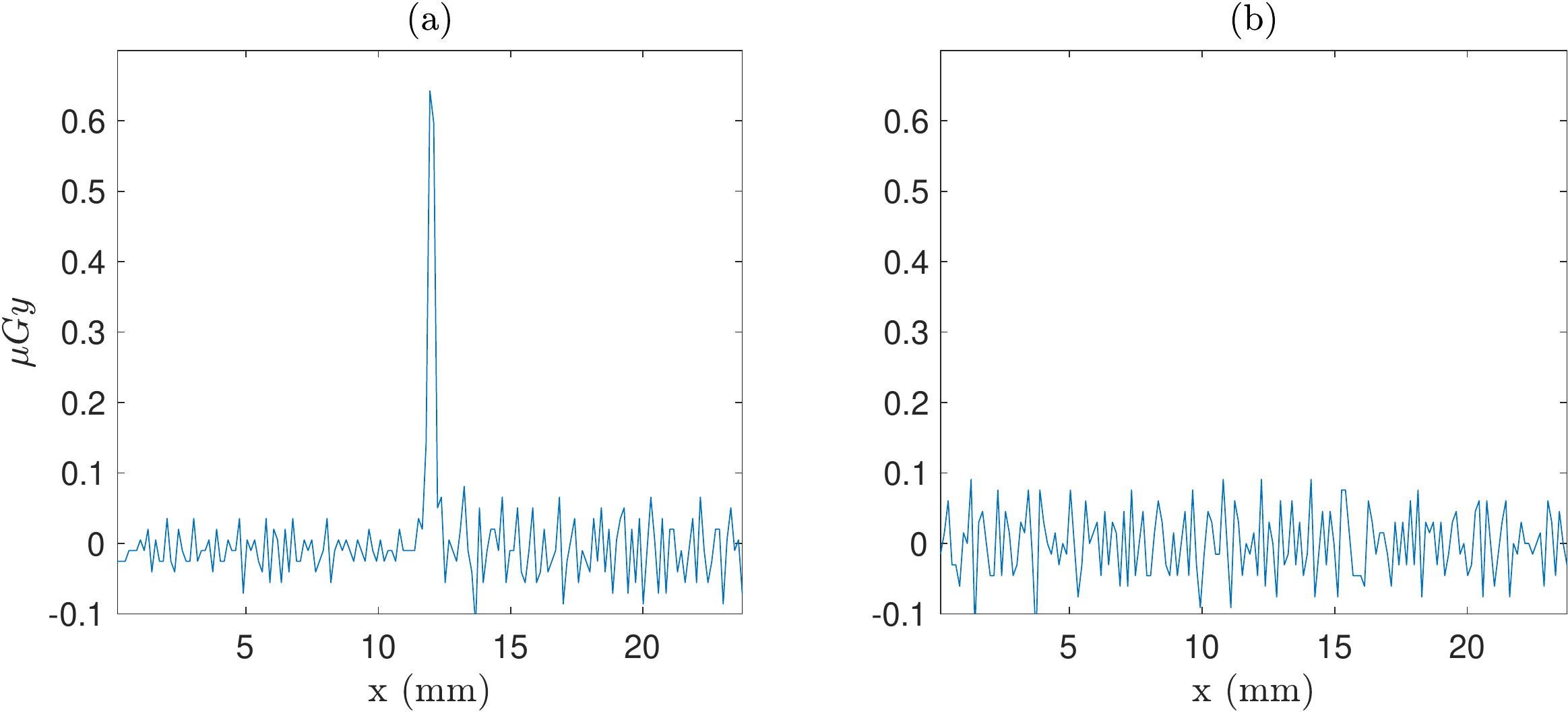}
\caption{Signals $g$ (a) and $n$ (b) produced as derivatives of profiles in figure \ref{fig:esf}.}
\label{fig:lsf}
\end{figure}

Figure \ref{fig:esf} shows two parts of the profile resulting from scanning the edge image in figure \ref{fig:edge} along the top row. The left side of figure \ref{fig:esf} show the central part of this edge profile whereas the right side shows the rightmost part of it. Signals to be detected are obtained from the numerical derivatives of these profiles (figure \ref{fig:lsf}).

In order to obtain the template $Hf$ in equation \ref{eq:d2npw} ($H\delta$ in this case), a noiseless synthetic image was generated by convolving an ideal edge image, with the same angle as the real edge, with the system's point spread function (PSF). The convolutions was carried out using a spatial resolution ten times higher than the sampling distance of the image and the PSF was obtained as the inverse DFT of the two-dimensional MTF of the system. After convolution, the resulting image was sampled at the image sampling distance and differentiated along the horizontal direction. Then, the template was calculated by aligning and averaging all the line spread profiles resulting from differentiation.

Signals used to assess detectability are obtained as high frequency residues (HFRs) of the template $H\delta(x)$ and HFRs of signals like $g(x)$ and $n(x)$ in figure \ref{fig:lsf}. In the digital system the spatial coordinate x takes values in $\{l\Delta\}_{l=0,..,N-1}$, where $N$ is the number of points in and $g(x)$ and $\Delta$ is the sampling distance in the image.

High frequency residues of order $k$, $r_{H\delta}^k(x)$, $r_g^k(x)$ and $r_n^k(x)$ with $k=1,2,..$ for template $H\delta(x)$, output $g(x)$ and noise $n(x)$ respectively are obtained in a three-step process. First, the discrete Fourier transform of the signal is calculated. Second all Fourier coefficients of order less than $k$ are made zero. Finally, the inverse transform is applied to the processed coefficients to obtain the residue. In this way, the lowest frequency component remaining in the residue of order $k$ corresponds to a frequency 
\begin{equation}\label{eq:frec}
u_k=\frac{k}{N\Delta}.
\end{equation}

\begin{figure*}
	\centering
	\includegraphics[width=140mm]{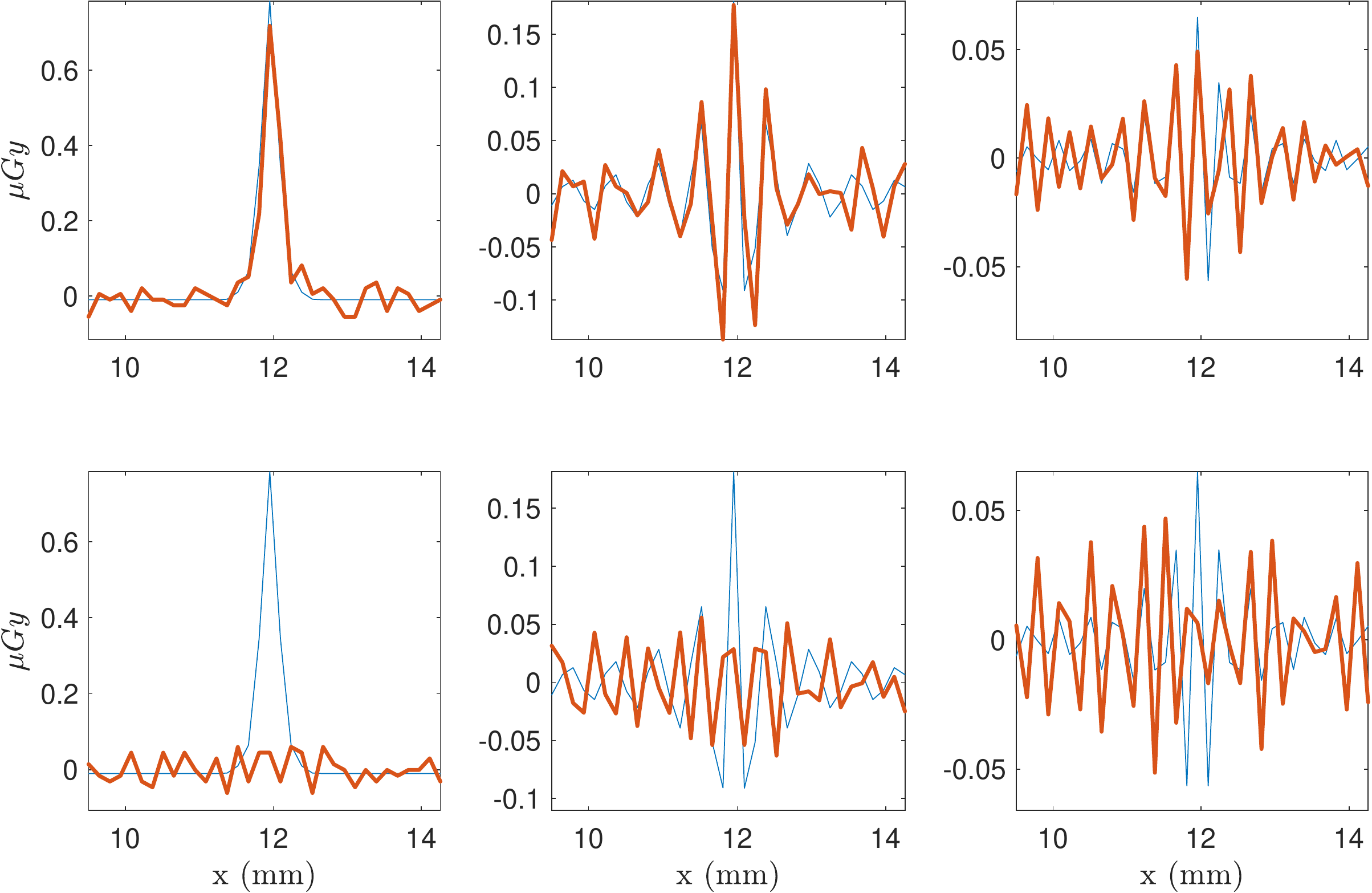}
	\caption{High frequency residues of orders $k=1$, $k=41$ and $k=61$ for the template $H\delta(x)$ (thin lines), for one of the outputs $g(x)$ (thick lines upper row) and one of the noise signals $n(x)$ (thick lines lower row).}
	\label{fig:residues}
\end{figure*}

\begin{figure*}
	\centering
	\includegraphics[width=140mm]{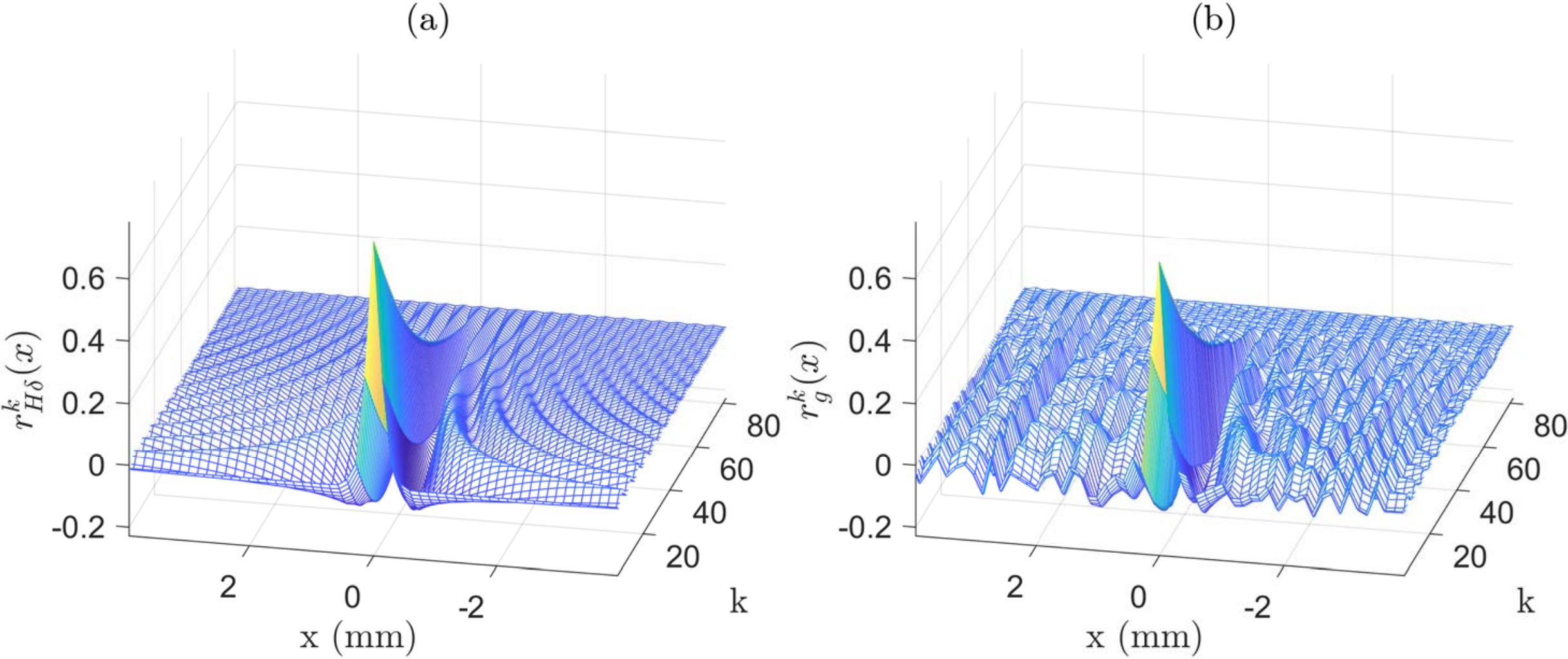}
	\caption{High frequency residues for the template $H\delta$ (a) and the output $g$ (b) as a function of the spatial coordinate $x$ for a beam quality RQA3 and an entrance air KERMA of 1.79 $\mu Gy$.}
	\label{fig:HFRs}
\end{figure*}

Figure \ref{fig:residues} shows the central part of high frequency residues (orders $k=1$, $k=41$ and $k=61$ from left to right) for the template $H\delta(x)$ (thin lines), for the output signal $g(x)$ (upper row and thick lines) and for the noise $n(x)$ (lower row and thick lines). It can be seen how, as the order of the HFR increases, the contrast and the low frequency contents of the residue are reduced. Both effects make signal $r_{H\delta}^k(x)$ harder to be detected as $k$ increases. Figure \ref{fig:HFRs} shows residues of template $H\delta(x)$ and output $g(x)$ for all possible values of $k$.

\begin{figure*}
	\includegraphics[width=140mm]{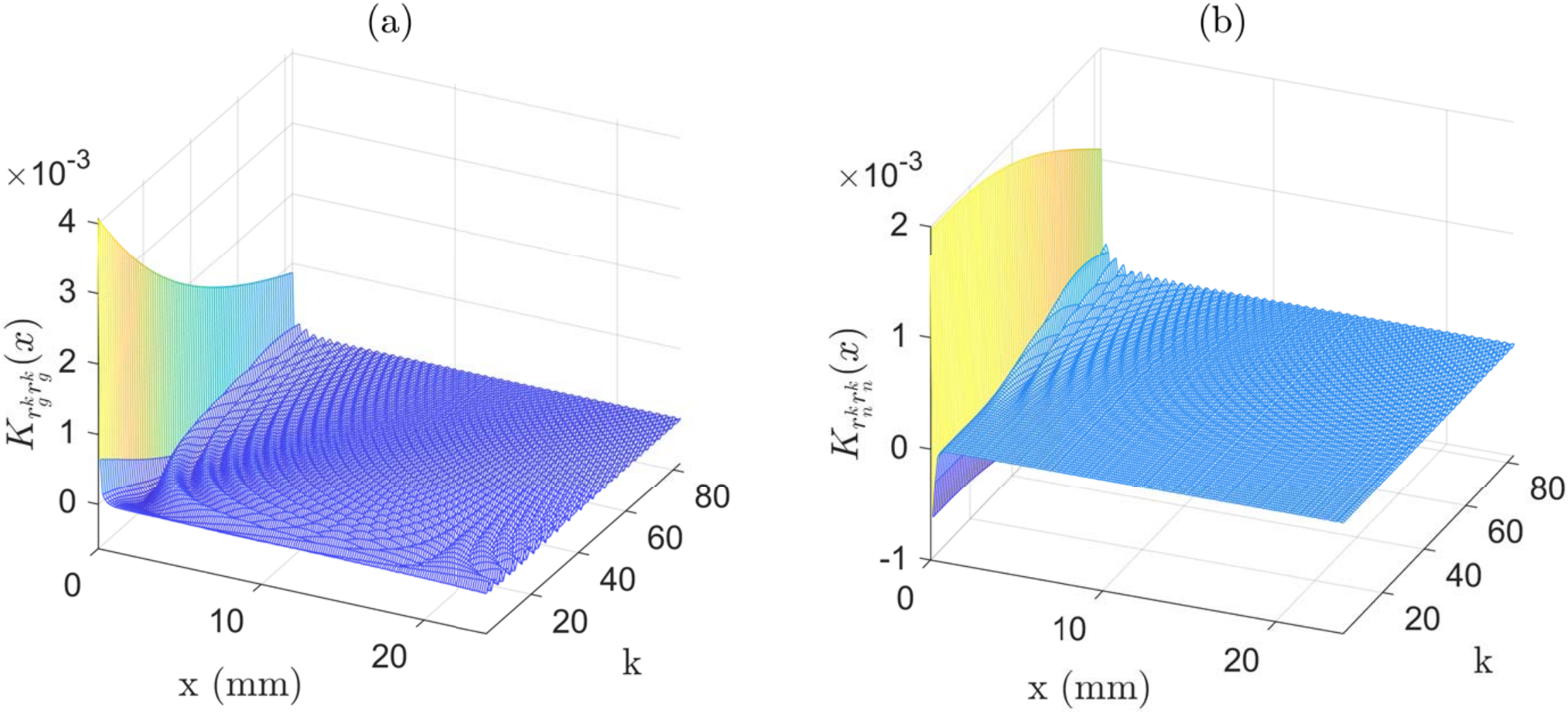}
\caption{Autocovariance functions for the different residues of order $k$ of output $g$ (a) and noise $n$ (b) for a beam quality RQA3 and an entrance air KERMA of 1.79 $\mu Gy$.}
\label{fig:Kxx}
\end{figure*}

Covariance matrices for residues $C_{r_g^k}$ and $C_{r_n^k}$ where obtained from the sample autocovariances \citep{10.5555/174458,gonzalez20b} $K_{r_g^kr_g^k}$ and $K_{r_n^kr_n^k}$ by $C_{r_g^k}(x,y)=K_{r_g^kr_g^k}(|x-y|)$ and $C_{r_n^k}(x,y)=K_{r_n^kr_n^k}(|x-y|)$ calculated using the profiles in 25 edge images per beam quality and dose. Figure \ref{fig:Kxx} shows the sample autocovariance for every residue of output $g=H\delta+n$ and noise $n$ as a function of the spatial coordinate $x$. It should be noted that noise $n$ is obtained by differentiating the image noise, which explains the negative values in $K_{r_n^0r_n^0}$.

NEQ was calculated following the IEC \cite{iec} as the quotient of the system squared MTF and the NNPS. In order to reduce uncertainties in the NEQ determination, the NNPS was calculated using an window mixing estimator \citep{gonzalez20b}.

\section{Results}
\begin{figure}
	\centering
		\includegraphics[width=80mm]{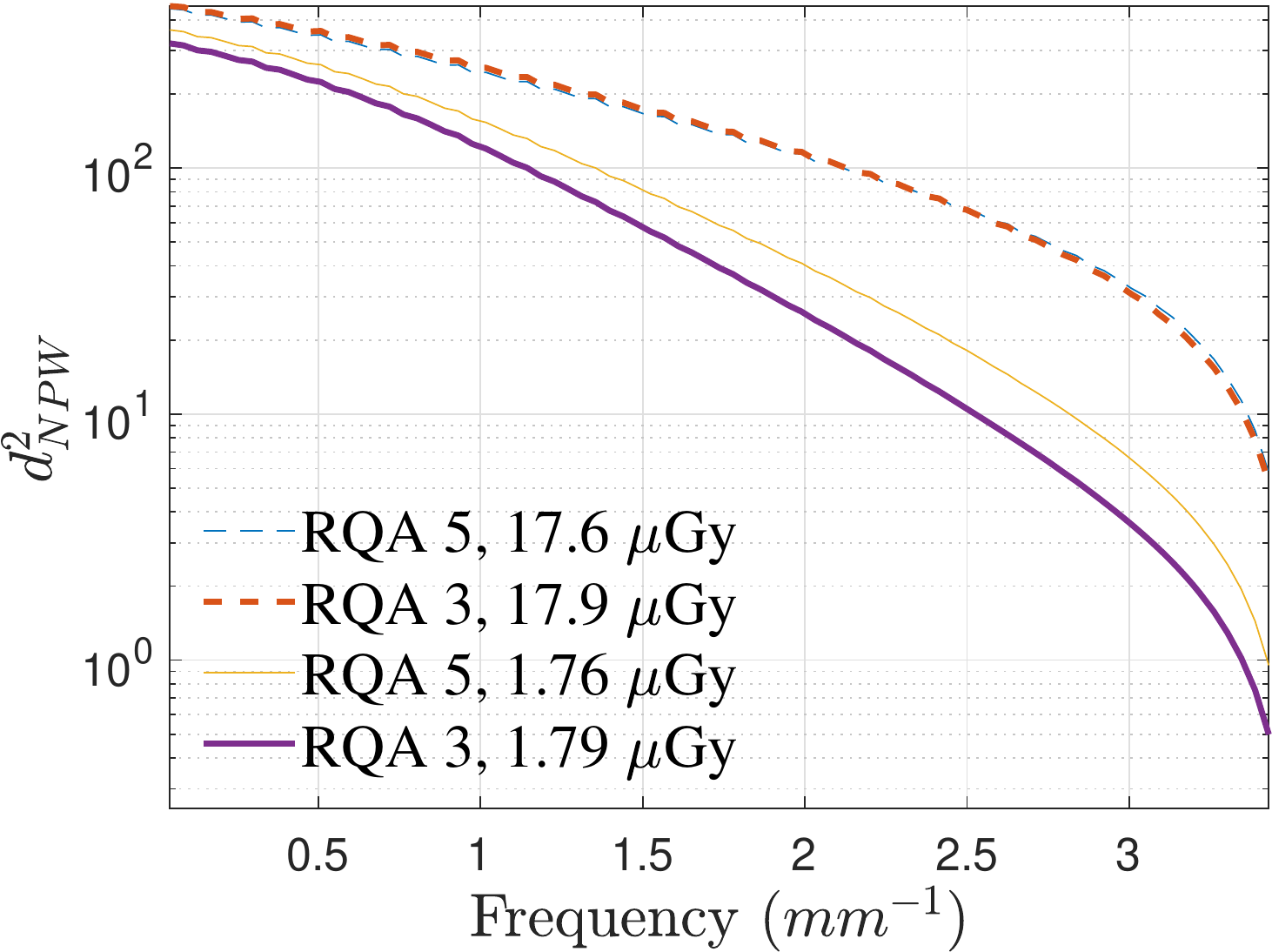}		
		\caption{Detectability indexes versus frequency for the different beam qualities and entrance air KERMAs used in this work.}
\label{fig:d2}
\end{figure}

Detectability indexes for the different beam qualities and entrance air KERMA studied at this work are shown in figure \ref{fig:d2}. For each combination, a curve presenting $d^2_{NPW}$ (equation \ref{eq:d2npw}) as a function of frequency is plotted. Here, each frequency value $u_k$ at the abscissa axis represents the frequency of the lowest frequency component remaining in residue $r_{H\delta}^k$ (equation \ref{eq:frec}). In order to evaluate $d^2_{NPW}$ at $u_k$, $Hf$ in equation \ref{eq:d2npw} is replaced by $r_{H\delta}^k$ and $C_g$ is replaced by $\frac{1}{2}(C_{r_{g}^k}+C_{r_n^k})$,

\begin{equation}\label{eq:d2large}
d^2_{NPW}(u_k)=\frac{((r_{H\delta}^k)^tr_{H\delta}^k)^2}{(r_{H\delta}^k)^t \frac{1}{2}(C_{r_{g}^k}+C_{r_n^k}) r_{H\delta}^k}.
\end{equation}

\begin{figure}
\centering
{\includegraphics[width=80mm]{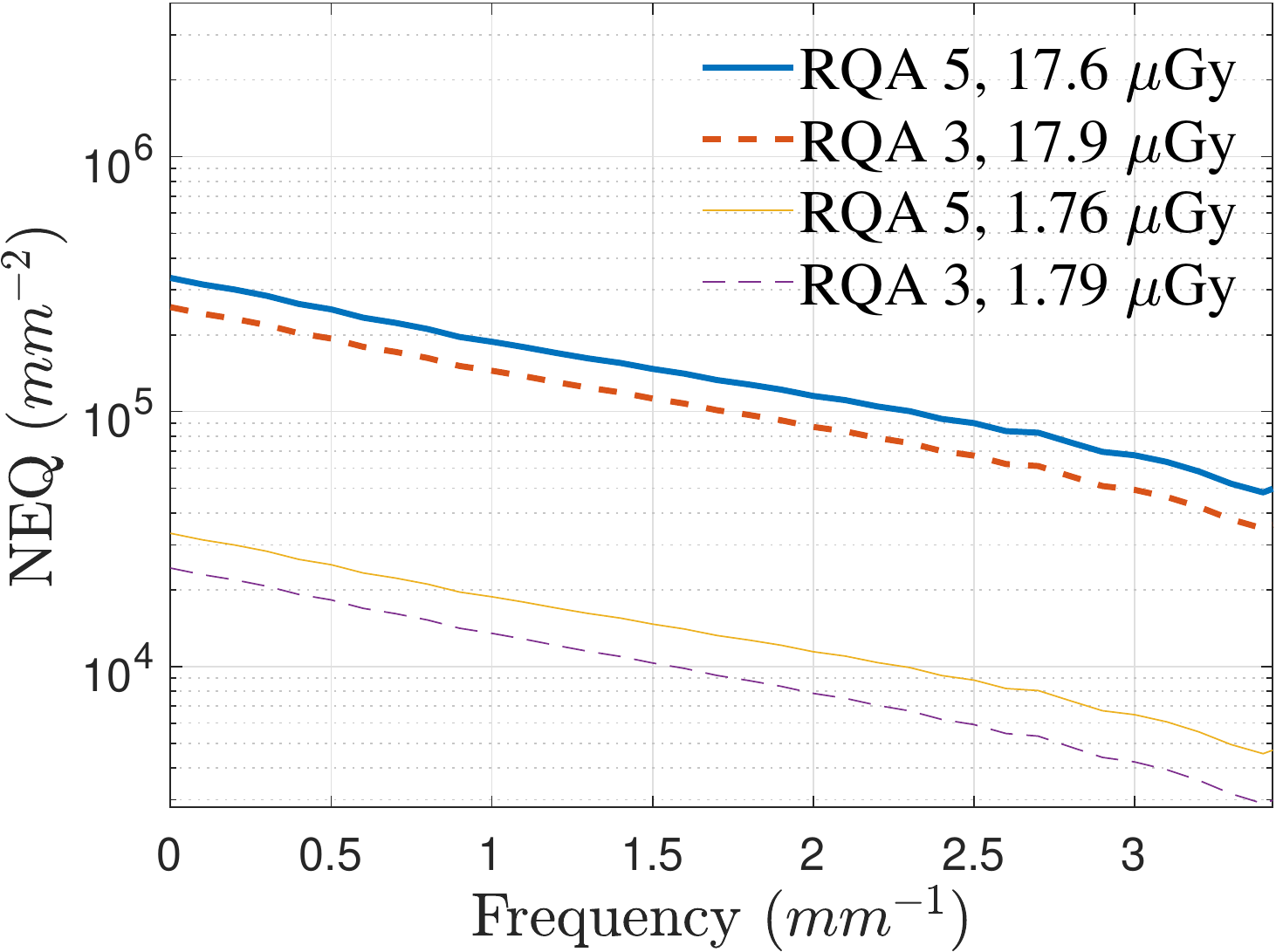}}
\caption{Noise equivalent quanta for the different techniques studied in this work.}
\label{fig:NEQ}
\end{figure}

It can be seen how $d^2_{NPW}$ is greater for RQA 5 than for RQA 3 at low doses. However, for the higher dose levels, $d^2_{NPW}$ takes similar values for both beam qualities. If these results are compared with those of NEQ (figure \ref{fig:NEQ}), the differences are notable since NEQ differences between beam qualities are similar at low and high doses.

\begin{figure}
	\centering
	\includegraphics[width=80mm]{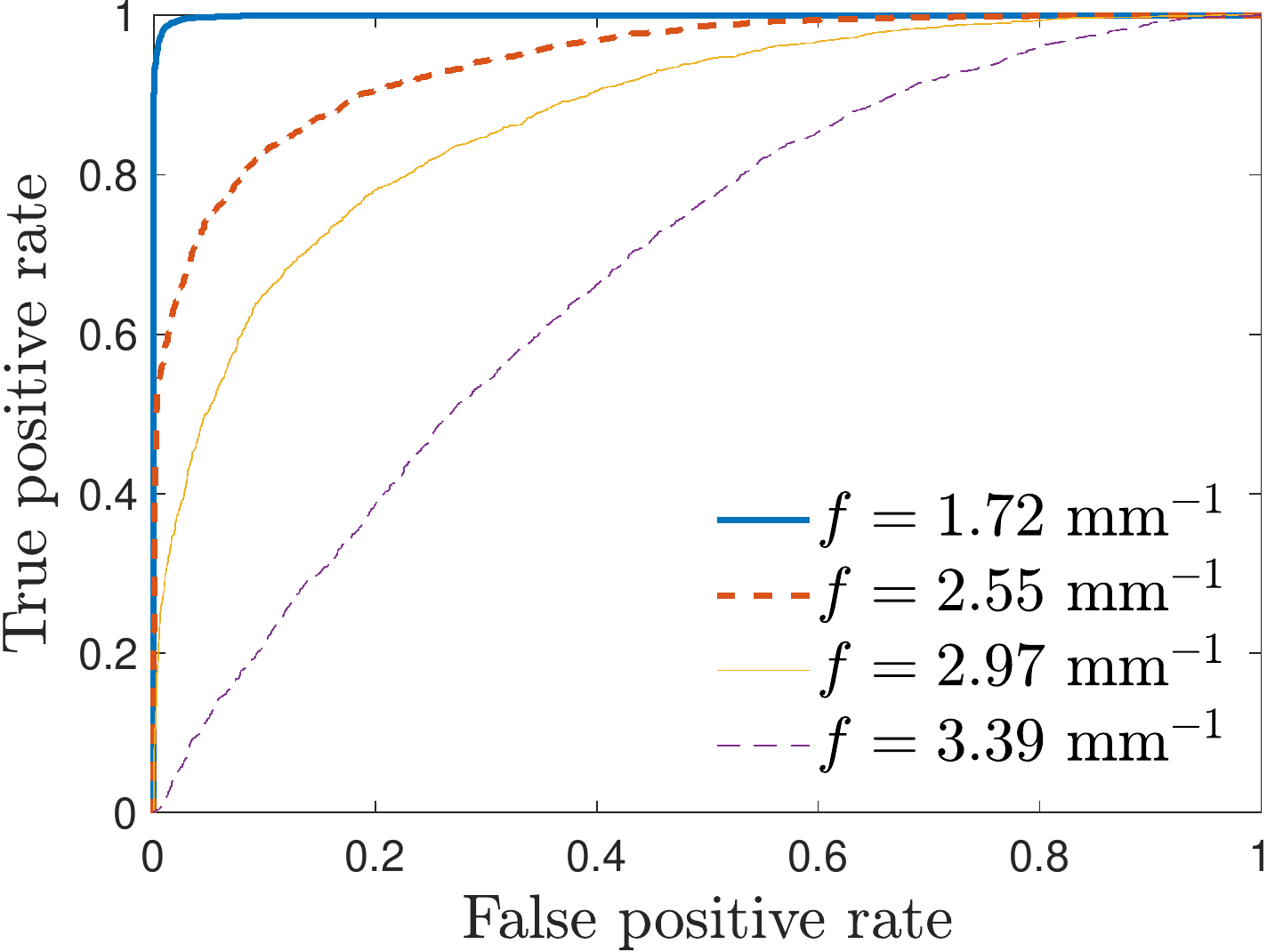}
	\caption{ROC curves for residues of order $k=41$, $k=61$, $k=71$ and $k=81$. The RQA 3 image acquired with 1.79 $\mu Gy$ of entrance air KERMA was used.}
	\label{fig:roc}
\end{figure}

In addition to the calculation of detectability indexes, ROC analysis was carried out using a NPW decision function, the template and all the different $g$ and $n$ profiles that could be extracted from the image in figure \ref{fig:edge}. Figure \ref{fig:roc} shows four of the ROC curves obtained. These curves correspond to residues $k=$ 41, 61, 71 and 81 of the top profile in figure \ref{fig:edge}, and these residues correspond to frequencies $u_{41}=1.72$, $u_{61}=2.55$, $u_{71}=2.97$, and $u_{81}=3.39$ mm$^{-1}$ ($N=166$ in equation \ref{eq:frec} for profiles of 23.9 mm length).

\begin{figure}
	\centering
		\includegraphics[width=80mm]{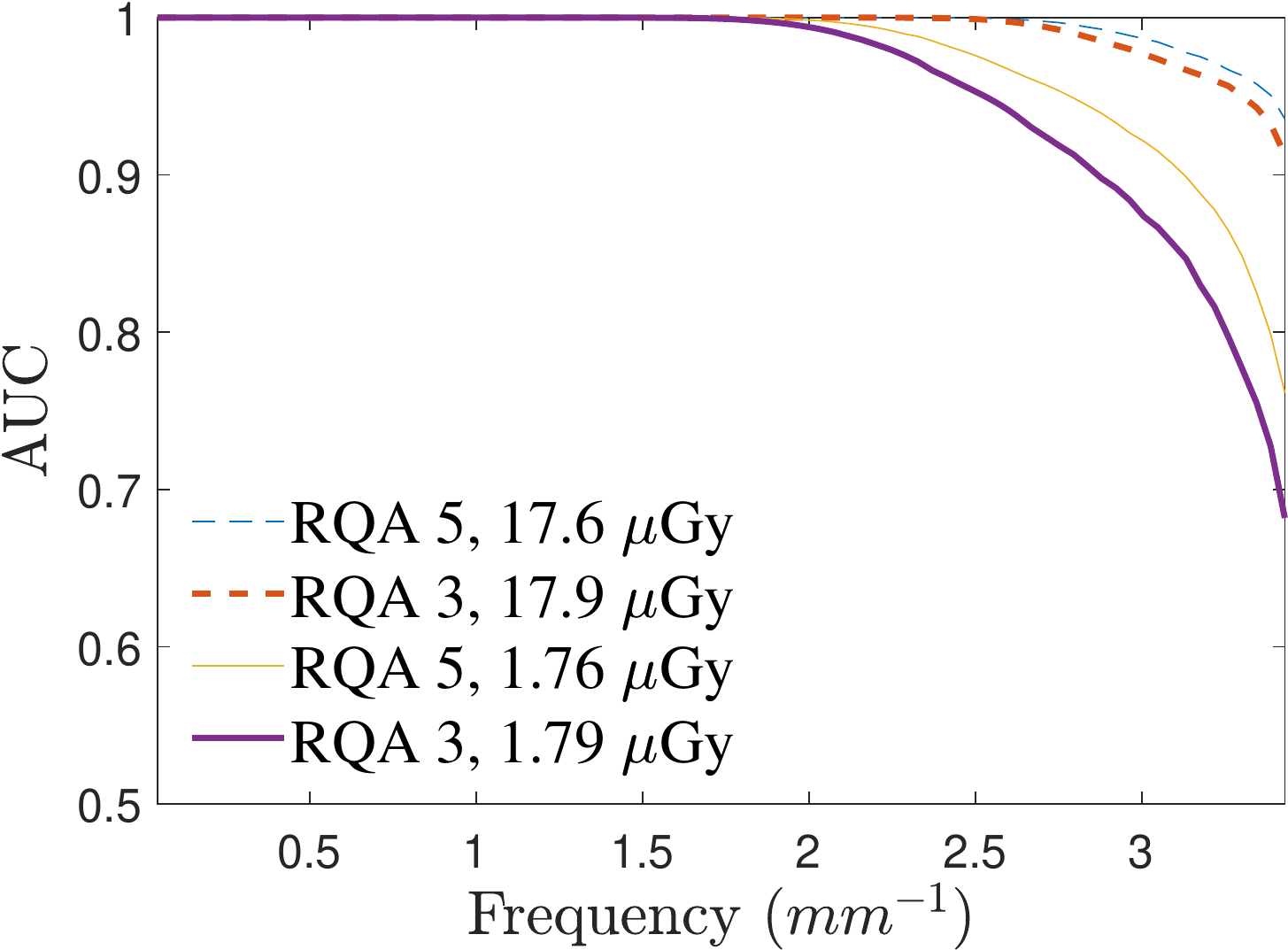}		
		\caption{Area under the ROC curve versus frequency for the different beam qualities and entrance air KERMAs used in this work.}
\label{fig:auc}
\end{figure}

Figure \ref{fig:auc} presents the area under the ROC curve for residue $k$ versus the frequency $u_k$. This figure shows the effect that dose and beam quality have in detectability of HFRs. The different acquisition techniques are classified in a similar way as $d^2_{NPW}$ did in figure \ref{fig:d2}. The most visible difference occurs at the higher doses, since in this case the two beam qualities are slightly distinguishable.

\section{Discussion}
The selection of HFR of line profiles to study detectability in an X-ray imaging system is not justified from a diagnostic task-based approach, since one-dimensional delta-type inputs are not the type of signal that is normally sought after in a diagnostic image. However, the characteristics of the signals presented in this work may be useful to investigate the capacity of the system for detecting signals of interest in diagnostic. From a mathematical point of view, the HFRs studied at this work represent the extreme opposite to the functions sought in the case of NEQ in terms of the support of their transform (the function used to generate the HFRs is a delta function and NEQ is detectability of a complex exponential function). For this reason, $d^2_{NPW}$ of HFRs can be seen as a complement to NEQ in the description of the system detectability performance. Also, the justification for using the higher and higher order residues of a delta function lies in that, in that way, we are studying detectability of those components in a delta that are more and more difficult to detect: we are looking for the limits of the system.

When comparing $d^2_{NPW}$ with NEQ (figures \ref{fig:d2} and \ref{fig:NEQ}), very different curves are obtained, which highlights the importance of the type of signal used for the study. The differences in detectability between qualities at different dose levels are noticeable. Contrary to $d^2_{NPW}$, NEQ differences between beam qualities are similar at low doses and at high doses. This was to be expected, since in the definition of NEQ (equation \ref{eq:neq}) the only difference between NEQ values for different beam qualities lies in the spectral density of the noise, and if we multiply the value of KERMA in both qualities by the same factor both densities will be modified in a similar way. In contrast, noise is represented in equation \ref{eq:d2large} by $C_{r_g^k}+C_{r_n^k}$ and this term tends to be independent of noise as the KERMA increases (it tends to $C_{r_g^k}$). This can also be seen in the AUC results (figure \ref{fig:auc}) where differences between beam qualities are much more important at low doses than at high doses.

Another aspect that calls attention between the results of $d^2_{NPW}$ and NEQ is how the index decreases with increasing frequency. The first method has a higher relative sensitivity (the same increase in frequency will produce a higher relative change in $d^2_{NPW}$ than in NEQ). Also, relative slopes for NEQ are similar for the different doses and beam qualities whereas for $d^2_{NPW}$ the change in slope with the dose is noticeable. Then, the relative sensitivity of $d^2_{NPW}$ decreases with increasing dose. 

All the differences found in this work between $d^2_{NPW}$ and NEQ stress the importance of both the signal used for the detectability task and the observer that has been implemented. A comprehensive study should take into account that detectability of an object in an image depends on physical properties of the imaging system like noise and spatial resolution, some characteristics of the object such as its shape, size and contrast and on the decision function of the observer model used \cite{ICRU1995,gang14}. Curves \ref{fig:d2} and \ref{fig:NEQ} incorporate measurements of noise and spatial resolution of the imaging system (NNPS and MTF in the definition of $d^2_{NPW}$ and NEQ); signals of very different shape and size (delta and harmonic functions for $d^2_{NPW}$ and NEQ respectiverly); different contrasts (contrast for residues $r_{H\delta}^k$ decreases with $k$) and two different observers.

\section{Conclusion}
Detectability assessment results of an imaging system depends on the type of signals used for the study. The selection of the signals should follow one of two criteria. On the one hand, a task-based approach could be implemented by using signals with a clinical relevance. This approach, however, could be too time consuming if a wide range of clinical applications are to be covered. On the other hand, using mathematical models covering a relatively large range of signals could provide an alternative evaluation. An example of these mathematical models is the set of complex exponential functions $f(x,y)=e^{- i2\pi (ux+vy)}$ for which detectability is described by the noise equivalent quanta. Since the spectral power of an exponential function is concentrated on a single frequency, the extreme opposite to this type of function is a generalized function delta $f(x,y)=\delta(x,y)$ for which the spectral density is uniformly distributed throughout the frequency space.

In this work, statistical decision theory has been applied to study the data acquisition stage in an x-ray imaging system for different beam qualities and different doses. The set of signals to be detected is built from a delta input and is obtained by progressively eliminating the lower frequency components of the delta. The resulting high frequency residues make up a series of signals in which contrast decreases as the order of the residue increases. At the same time, the frequency band of the residues is shifted to higher frequencies. The fact that both effects make detectability of the residue more difficult can been exploited to investigate the limits of the system performance. 

The information provided by detectability of these HFRs is a complement to that of the NEQ, since both indexes uses functions with a very different representation in terms of spatial support and frequency content.

Finally, it has been discussed how the particular observer selected determines the final results. The conclusion that is obtained from the NPW observer is that at high doses detectability does not depend on the quality of the beam, contrary to what is inferred from the IBO observer used in the calculation of NEQ. 

\appendix
\section{Calculation of the detectability index}\label{ap:1}
In the case of signal and background known exactly (SKE/BKE), for a discrete linear system and Gaussian noise, the detectability index $d^2_{IBO}$ or squared SNR of the decision function $L$ of an ideal Bayesian observer is expressed as the product of the matrices \cite{ICRU1995,fukunaga90}
\begin{equation}\label{eq_decfun}
d^2_{IBO}=f^tH^tC^{-1}_nHf,
\end{equation}
where $f$ is the imaged object, the superscript $^t$ stands for the transpose of a matrix, $H$ is the linear system transfer function and $C_n$ is the covariance matrix of noise.

By inserting the identity matrix $I=W^*W$ in equation \ref{eq_decfun}, where $W$ is the matrix defining the unitary DFT with matrix elements 
\begin{equation}
W_{k,l}=\frac{e^{-i2\pi kl/N}}{\sqrt{N}},\:k,l=0,..N-1, 
\end{equation}
$N$ is the length of the signals and the asterisk denotes the matrix with complex conjugated entries (not the Hermitian matrix), $d^2_{IBO}$ can be rewritten as
\begin{equation}\label{eq:step1}
d^2_{IBO}=f^tH^tW^*WC^{-1}_nHf.
\end{equation}

If we consider a periodic and shift invariant system, the matrix $H$ will be a circulant matrix and its autovectors and autovalues will be the columns of $W$ and the components of the system OTF $\hat{h}$ (the DFT of the first row $h$ of $H$) respectively. Therefore
\begin{equation}
HW=Wdiag(\hat{h}),
\end{equation}  
where $diag(v)$ is the diagonal matrix in which the diagonal elements are the elements of vector $v$. Using this result and assuming that $H$ is real and symmetric the term $H^tW^*$ in equation \ref{eq:step1} can be written as
\begin{equation}
H^tW^*=HW^*=(HW)^*=(Wdiag(\hat{h}))^*=W^* diag(\hat{h}^*).
\end{equation}  
In this way, since $f^tW^*=\hat{f}^{*t}$ equation \ref{eq:step1} can be written as
\begin{equation}
d^2_{IBO}=\hat{f}^{*t}diag(\hat{h}^*)WC^{-1}_nHf.
\end{equation}

If noise is wide sense stationary, $C_n$ is circulant. Since it is also real and symmetric then $C_nW=Wdiag(\hat{c})$, with $\hat{c}$ being the DFT of the first row of $C_n$. Then
\begin{equation}
C_n=W^*WC_n=W^*(Wdiag(\hat{c}))^t=W^*diag(\hat{c})W
\end{equation} 
and
\begin{equation}
C_n^{-1}=W^*diag(\hat{c})^{-1}W.
\end{equation}

Therefore, equation \ref{eq_decfun} can be written as
\begin{equation}
d^2_{IBO}=\hat{f}^{t*}diag(\hat{h}^*)diag(\hat{c})^{-1}diag(\hat{h})\hat{f},
\end{equation}
or as
\begin{equation}\label{eq:casifinal}
d^2_{IBO}=\sum_{l=0}^{N-1}\frac{\hat{f}^*_l \hat{h}^*_l   \hat{h}_l  \hat{f}_l}{\hat{c}_l}=
\sum_{l=0}^{N-1}\frac{|\hat{f}_l|^2 |\hat{h}_l|^2}{\hat{c}_l}.
\end{equation}



\end{document}